\documentclass[12pt,a4wide,epsf]{article}
\usepackage{epsf}

\usepackage{graphicx}

\newcommand{\insertplot}[5]{\begin{figure}
 \hfill\hbox to 0.05in{\vbox to #5in{\vfill
 \inputplot{#1}{#4}{#5}}\hfill}
 \hfill\vspace{-.1in}
 \caption{#2}\label{#3}
 \end{figure}}
 \newcommand{\inputplot}[3]{
 \special{ps: plotfile #1}
}
\newcounter{fig}   

\newcommand{\vphi}{\varphi}
\newcommand{\vepsilon}{\varepsilon}

\newcommand{\sqdetg}{\sqrt{-g}}

\usepackage{graphicx}

\begin{document}

\title{
Charged Rotating Black Holes\\ in Odd Dimensions}
 \vspace{1.5truecm}
\author{
{\bf Jutta Kunz},
%{\bf Jutta Kunz} and
{\bf Francisco Navarro-L\'erida} and
%{\bf Francisco Navarro-L\'erida}\\
{\bf Jan Viebahn}\\
Institut f\"ur Physik, Universit\"at Oldenburg, Postfach 2503\\
D-26111 Oldenburg, Germany\\
}

\vspace{1.5truecm}

%\date{April 14, 2006}
\date{\today}

\maketitle
\vspace{1.0truecm}

\begin{abstract}
We consider charged rotating black holes 
%in Einstein-Maxwell theory 
in $D=2N+1$ dimensions, $D \ge 5$.
%which are asymptotically flat
%and possess a regular horizon of spherical topology.
While these black holes generically possess $N$ independent angular momenta,
associated with $N$ distinct planes of rotation,
we here focus on black holes with equal-magnitude angular momenta.
The angular dependence can then be treated explicitly,
and a system of 5 $D$-dependent ordinary differential equations is obtained.
We solve these equations numerically for Einstein-Maxwell theory 
in $D=5$, 7 and 9 dimensions.
We discuss the global and horizon properties of these black holes,
as well as their extremal limits.
\end{abstract}

\vfill\eject

\section{Introduction}

In $D=4$ dimensions the Kerr-Newman solution presents 
the unique family of stationary asymptotically flat black holes
of Einstein-Maxwell (EM) theory.
It comprises the Kerr solution,
representing rotating vacuum black holes,
as well as the static Reissner-Nordstr\"om and Schwarzschild solutions.

The generalization of these black hole solutions to $D>4$
dimensions was pioneered by Tangherlini
\cite{tangher} for static black holes,
and by Myers and Perry \cite{MP} for rotating vacuum black holes.
The corresponding $D>4$ charged rotating black holes of EM theory 
could not yet be obtained in closed form \cite{MP,Horo}.
But in $D=5$ dimensions rotating EM black holes have been constructed 
numerically \cite{KNP}.
%for a single angular momentum or two equal-magnitude

In contrast to pure EM theory, exact higher dimensional
charged rotating black holes are known in theories with more symmetries.
The presence of a Chern-Simons (CS) term, for instance, 
leads to a class of odd-dimensional
Einstein-Maxwell-Chern-Simons (EMCS) theories,
comprising the bosonic sector of minimal $D=5$ supergravity,
whose stationary black hole solutions \cite{BLMPSV,BMPV,Cvetic}
possess surprising properties \cite{GMT,surprise,KN}.
In particular, EMCS black holes (with horizons of spherical topology)
are no longer uniquely characterized by their global charges \cite{KN}.
The inclusion of additional fields, as required by supersymmetry
or string theory, yields further exact solutions \cite{Cvetic7,Youm,Horo2}.

Stationary black holes in $D$ dimensions 
possess $N=[(D-1)/2]$ independent angular momenta $J_i$ 
associated with $N$ orthogonal planes of rotation \cite{MP}.
($[(D-1)/2]$ denotes the integer part of $(D-1)/2$, corresponding to the
rank of the rotation group $SO(D-1)$.) 
The general black holes solutions then fall into two classes,
in even-$D$ and odd-$D$ solutions \cite{MP}.

Here we focus on charged rotating black holes in odd dimensions. We show, that
when their $N$ angular momenta have all equal magnitude, 
the angular dependence can be treated explicitly
%,thus generalizing this observation from $D=5$ to arbitrary
for any odd dimension $D \ge 5$.
% \cite{BMPV,string2}.
The resulting system of coupled Einstein and matter field equations 
then simplifies considerably,
yielding a system of $D$-dependent ordinary differential equations.
We here solve these equations numerically for Einstein-Maxwell theory
in $D=5$, 7 and 9 dimensions.

In section 2 we recall the EM action, 
and present the stationary axially symmetric Ans\"atze
for black hole solutions with $N$ equal-magnitude angular momenta
in $D=2N+1$ dimensions, $N\ge 2$.
We discuss the black hole properties in section 3,
and present numerical results for EM black holes in section 4.

\section{Metric and Gauge Potential}

We consider the $D$-dimensional Einstein-Maxwell action
with Lagrangian
\begin{equation}
L = \frac{1}{16 \pi G_D} \sqdetg  (R - F_{\mu \nu} F^{\mu \nu}) \ ,
\end{equation}
with curvature scalar $R$,
$D$-dimensional Newton constant $G_D$,
and field strength tensor
$
F_{\mu \nu} =
\partial_\mu A_\nu -\partial_\nu A_\mu $,
where $A_\mu $ denotes the gauge potential.

Variation of the action with respect to the metric and the gauge potential
leads to the Einstein equations 
\begin{equation}
G_{\mu\nu}= R_{\mu\nu}-\frac{1}{2}g_{\mu\nu}R = 2 T_{\mu\nu}
\ , \label{ee}
\end{equation}
with stress-energy tensor
\begin{equation}
T_{\mu \nu} = F_{\mu\rho} {F_\nu}^\rho - \frac{1}{4} g_{\mu \nu} F_{\rho
  \sigma} F^{\rho \sigma} \ ,
\end{equation}
and the gauge field equations,
\begin{equation}
\nabla_\mu F^{\mu\nu}  = 0 \ .
\label{feqA}
\end{equation}

To obtain stationary black hole solutions,
representing charged generalizations of the D-dimensional
Myers-Perry solutions \cite{MP},
we consider black hole space-times with $N$-azimuthal symmetries,
implying the existence of $N+1$ commuting Killing vectors,
$\xi \equiv \partial_t$, 
and $\eta_{(k)} \equiv \partial_{\vphi_k}$, for $k=1, \dots , N$.
We parametrize the metric in isotropic coordinates,
which are well suited for the numerical construction of
rotating black holes \cite{KNP,KN,kkrot}.
(We consider only black holes with spherical horizon topology
\cite{blackrings}.)

While the general EM black holes will then possess $N$ independent
angular momenta, we now restrict to black holes whose 
angular momenta have all equal magnitude.
The metric and the gauge field parametrization then
simplify considerably. 
In particular, for such equal-magnitude angular momenta
black holes, the general Einstein and Maxwell equations reduce to
a set of ordinary differential equations.

The metric for these equal-magnitude angular momenta 
black holes reads
\begin{eqnarray}
&&ds^2 = -f dt^2 + \frac{m}{f} \left[ dr^2 + r^2 \sum_{i=1}^{N-1}
  \left(\prod_{j=0}^{i-1} \cos^2\theta_j \right) d\theta_i^2\right] \nonumber \\
&&+\frac{n}{f} r^2 \sum_{k=1}^N \left( \prod_{l=0}^{k-1} \cos^2 \theta_l
  \right) \sin^2\theta_k \left(\vepsilon_k d\vphi_k - \frac{\omega}{r}
  dt\right)^2 \nonumber \\
&&+\frac{m-n}{f} r^2 \left\{ \sum_{k=1}^N \left( \prod_{l=0}^{k-1} \cos^2
  \theta_l \right) \sin^2\theta_k  d\vphi_k^2 \right. \nonumber\\
&& -\left. \left[\sum_{k=1}^N \left( \prod_{l=0}^{k-1} \cos^2
  \theta_l \right) \sin^2\theta_k \vepsilon_k d\vphi_k\right]^2 \right\} \ ,
\end{eqnarray}
where $\theta_0 \equiv 0$, $\theta_i \in [0,\pi/2]$ 
for $i=1,\dots , N-1$, 
$\theta_N \equiv \pi/2$, $\vphi_k \in [0,2\pi]$ for $k=1,\dots , N$,
and $\vepsilon_k = \pm 1$ denotes the sense of rotation
in the $k$-th orthogonal plane of rotation.

An adequate parametrization for the gauge potential is given by
\begin{equation}
A_\mu dx^\mu =  a_0 dt + a_\vphi \sum_{k=1}^N \left(\prod_{l=0}^{k-1}
  \cos^2\theta_l\right) \sin^2\theta_k \vepsilon_k d\vphi_k \ .
\end{equation}
Thus, independent of the odd dimension $D\ge 5$,
this parametrization involves only four functions $f, m, n, \omega$
for the metric and two functions $a_0, a_\vphi$
for the gauge field, which all depend only on the radial coordinate $r$.

To obtain asymptotically flat solutions, 
the metric functions should satisfy
at infinity the boundary conditions
\begin{equation}
f|_{r=\infty}=m|_{r=\infty}=n|_{r=\infty}=1 \ , \ \omega|_{r=\infty}=0 \ ,
\label{bc1} \end{equation}
while for the gauge potential we choose a gauge, in which it vanishes
at infinity
\begin{equation}
a_0|_{r=\infty}=a_\vphi|_{r=\infty}=0 \ .
\label{bc2} \end{equation}

The horizon is located at $r_{\rm H}$,
and is characterized by the condition $f(r_{\rm H})=0$ \cite{kkrot}.
Requiring the horizon to be regular, the metric functions must
satisfy the boundary conditions
\begin{equation}
f|_{r=r_{\rm H}}=m|_{r=r_{\rm H}}=n|_{r=r_{\rm H}}=0 \ ,
\ \omega|_{r=r_{\rm H}}=r_{\rm H} \Omega \ , 
\label{bc3} \end{equation}
where $\Omega$ is (related to) the horizon angular velocity, 
defined in terms of the Killing vector
\begin{equation}
\chi = \xi + \Omega \sum_{k=1}^N \vepsilon_k \eta_{(k)} \ ,
\label{chi} \end{equation}
which is null at the horizon. Without loss of generality, $\Omega$ is assumed
to be non-negative, any negative sign being included in $\vepsilon_k$.
The gauge potential satisfies
\begin{equation}
\left. \chi^\mu A_\mu \right|_{r=r_{\rm H}} =
\Phi_{\rm H} = \left. (a_0+\Omega a_\vphi)\right|_{r=r_{\rm H}} \ , \ \ \
\left. \frac{d a_\vphi}{d r}\right|_{r=r_{\rm H}}=0
\ , \label{bc4} \end{equation}
with constant horizon electrostatic potential $\Phi_{\rm H}$.

\section{Black Hole Properties}

The mass $M$ and the angular momenta $J_{(k)}$ of the black holes 
are obtained from the Komar expressions 
associated with the respective Killing vector fields
\begin{equation}
M = \frac{-1}{16 \pi G_D} \frac{D-2}{D-3} \int_{S_{\infty}^{D-2}} \alpha \ , \ \ \
J_{(k)} = \frac{1}{16 \pi G_D}  \int_{S_{\infty}^{D-2}} \beta_{(k)} \ ,
\end{equation}
with $\alpha_{\mu_1 \dots \mu_{D-2}} \equiv \epsilon_{\mu_1 \dots \mu_{D-2}
  \rho \sigma} \nabla^\rho \xi^\sigma$,
$\beta_{ (k) \mu_1 \dots \mu_{D-2}} \equiv \epsilon_{\mu_1 \dots \mu_{D-2}
  \rho \sigma} \nabla^\rho \eta_{(k)}^\sigma$,
and for equal-magnitude angular momenta $|J_{(k)}|=J$, 
$k=1, \dots , N$.

The electric charge is obtained from
\begin{equation}
Q=\frac{-1}{8 \pi G_D} \int_{S_{\infty}^{D-2}} \tilde F \ ,
\end{equation}
with ${\tilde F}_{\mu_1 \dots \mu_{D-2}} \equiv  \epsilon_{\mu_1 \dots \mu_{D-2} \rho \sigma} F^{\rho \sigma}$.

The horizon mass $M_{\rm H}$ and horizon angular momenta
$J_{{\rm H} (k)}$ are given by
\begin{equation}
M_{\rm H} = \frac{-1}{16 \pi G_D} \frac{D-2}{D-3} \int_{{\cal H}} \alpha \ , \ \ \ 
J_{{\rm H} (k)} = \frac{1}{16 \pi G_D}  \int_{{\cal H}} \beta_{(k)} \ ,
\end{equation}
where ${\cal H}$ represents the surface of the horizon,
and for equal-magnitude angular momenta 
$|J_{{\rm H} (k)}| =J_{\rm H}$, $k=1, \dots , N$.

Introducing further the area of the horizon $A_{\rm H}$ and
the surface gravity $\kappa$,
\begin{equation}
\kappa^2 =-\frac{1}{2} (\nabla_\mu \chi_\nu) (\nabla^\mu \chi^\nu) \ ,
\end{equation}
the mass formulae \cite{KNP,GMT} for EM black holes
with $N$ equal-magnitude angular momenta become
\begin{equation}
\frac{D-3}{D-2} M_{\rm  H} = \frac{\kappa A_{\rm H}}{8 \pi G_D} + N \Omega
J_{\rm H} \ ,
\end{equation}
\begin{equation}
\frac{D-3}{D-2} M = \frac{\kappa A_{\rm H}}{8 \pi G_D} + N \Omega
J  + \frac{D-3}{D-2} \Phi_{\rm H} Q  \ . \label{smarr}
\end{equation}

The global charges and the magnetic moment $\mu_{\rm mag}$,
can be obtained from the asympotic expansions of the metric and the gauge
potential 
\begin{eqnarray}
f=1-\frac{\hat M}{r^{D-3}} + \dots \ ,  \ \ \
\omega=\frac{\hat J}{r^{D-2}}  + \dots \ ,  \ \ \
a_0=\frac{\hat Q}{r^{D-3}} + \dots \ ,  \ \ \
a_\vphi=-\frac{{\hat \mu}_{\rm mag}}{r^{D-3}} + \dots \ ,
\end{eqnarray}
where
\begin{eqnarray}
{\hat M}=\frac{16 \pi G_D}{(D-2)A(S^{D-2})} M \ &,& \ \ \
{\hat J}=\frac{8 \pi G_D}{A(S^{D-2})}J \ , \nonumber \\
{\hat Q}=\frac{4 \pi G_D}{(D-3)A(S^{D-2})} Q \ &,& \ \ \
{\hat \mu}_{\rm  mag}=\frac{4 \pi G_D}{(D-3)A(S^{D-2})} \mu_{\rm mag} \ ,
\end{eqnarray}
and $A(S^{D-2})$ is the area of the unit $(D-2)$-sphere.
The gyromagnetic ratio $g$ is defined via
\begin{equation}
{\mu_{\rm mag}}=g \frac{Q J}{2M}
\ . \end{equation}

\section{Numerical results}

In order to solve the coupled system of ODE's, 
we take advantage of the existence of a
first integral of that system, 
\begin{equation}
\frac{r^{D-2} m^{(D-5)/2}}{f^{(D-3)/2}} \sqrt{\frac{m n}{f}} \left(\frac{d
      a_0}{dr} + \frac{\omega}{r} \frac{d a_\vphi}{dr} \right) = - \frac{4 \pi
      G_D}{A(S^{D-2})} Q \ ,
\end{equation}
to eliminate $a_0$ from the equations, leaving a 
system of one first order equation (for $n$) and
four second order equations.

For the numerical calculations we introduce
the compactified radial coordinate
$\bar{r}= 1-r_{\rm H}/r$ \cite{kkrot}, and we take units such that $G_D=1$.
We employ a collocation method for boundary-value ordinary
differential equations, equipped with an adaptive mesh selection procedure
\cite{COLSYS}.
Typical mesh sizes include $10^3-10^4$ points.
The solutions have a relative accuracy of $10^{-10}$.
The estimates of the relative errors of the global charges
and the magnetic moment are of order $10^{-6}$, 
giving rise to an estimate of the relative error of $g$ of order $10^{-5}$. 

Let us first address the domain of existence
of rotating EM black holes with equal-magnitude
angular momenta.
We note, that unlike the case of a single non-vanishing angular momentum,
where no extremal solutions exist in $D>5$ dimensions \cite{MP,Horo2},
extremal solutions do exist for odd $D$ black holes with
equal-magnitude angular momenta.
We exhibit in Fig.~1
the scaled angular momentum $J/M^{(D-2)/(D-3)}$ 
of the extremal EM black holes
versus the scaled charge $Q/M$ \cite{foot1}
for $D=5$, 7 and 9 dimensions.
Black holes exist only in the regions bounded by the
$J=0$-axis and by the respective curves.
The domain of existence is symmetric with respect to $Q \rightarrow -Q$.
Introducing the scaling factors $\delta$ and $\gamma$,
\begin{equation}
\delta^2=\frac{1}{2}\frac{D-2}{D-3} \ , \ \ \ (2\gamma)^{D-3}=\frac{1}{32\pi}
(D-3) (D-2)^{D-2} \left(\frac{D-1}{D-3}\right)^{(D-1)/2} A(S^{D-2}) \ ,
\end{equation}
we observe that the scaled domain of existence  
becomes almost independent of $D$, as demonstrated in Fig.~1 (right). 
These extremal black holes have vanishing surface gravity,
but finite horizon area.

\begin{figure}[h!]
\parbox{\textwidth}
{\centerline{
\mbox{
\epsfysize=10.0cm
\includegraphics[width=70mm,angle=0,keepaspectratio]{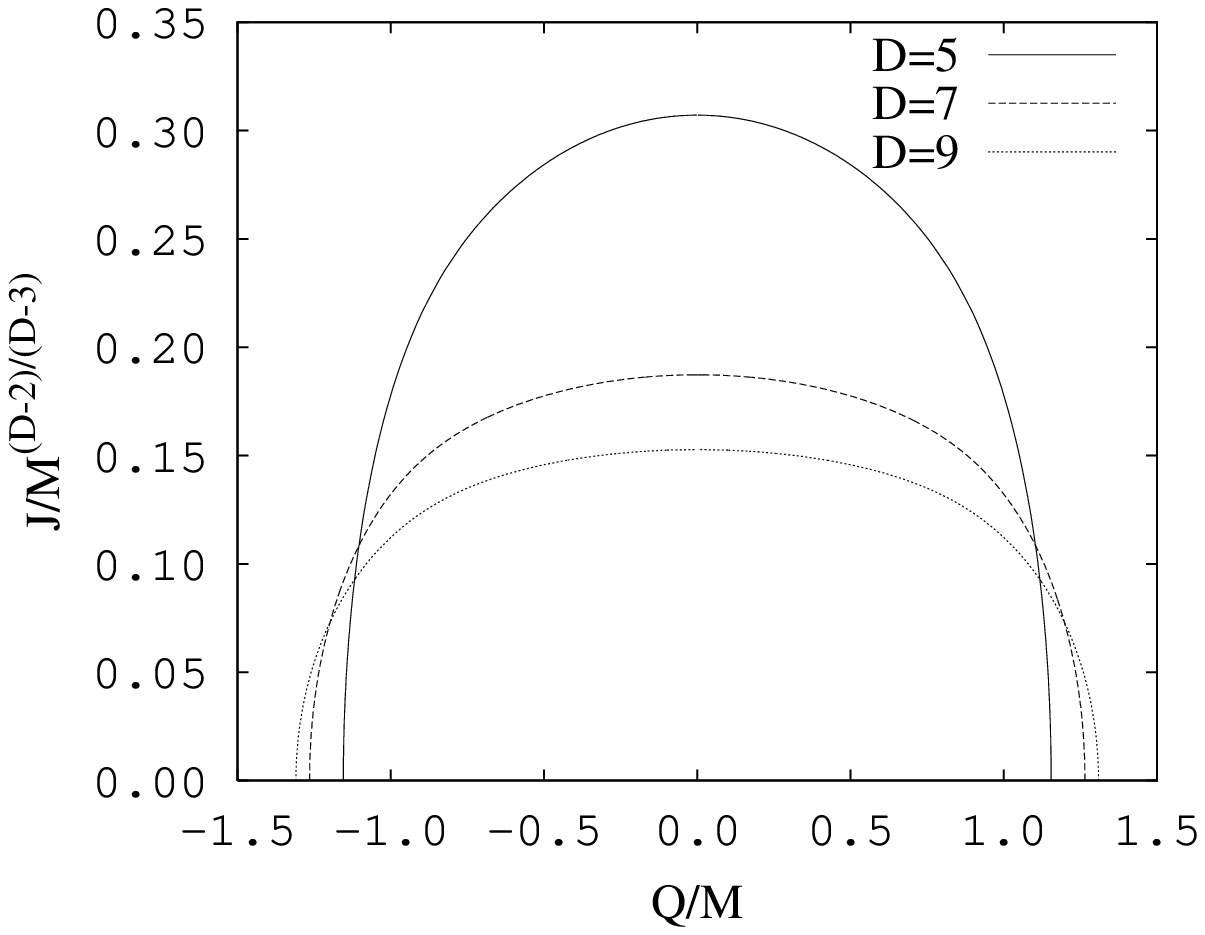}
\includegraphics[width=70mm,angle=0,keepaspectratio]{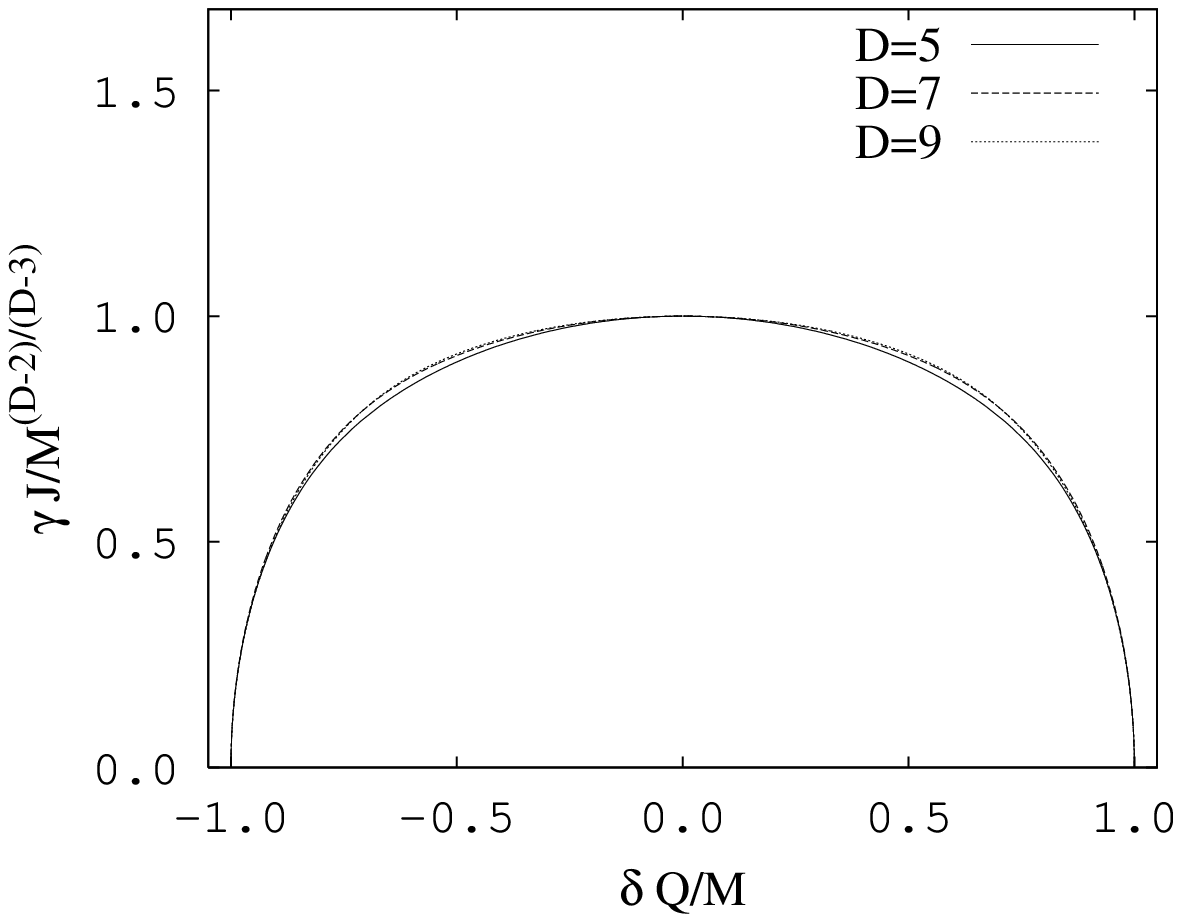}
}}}
\caption{
Left: Scaled angular momentum $J/M^{(D-2)/(D-3)}$ versus
scaled charge $Q/M$ for extremal black holes 
with equal-magnitude angular momenta
in $D=5$, 7 and 9 dimensions.
Right: Scaled domain of existence.
%Also indicated are the sets of 
%non-extremal solutions, exhibited in Figs.~2-4 (right).
}
\end{figure}

We now turn to non-extremal black holes,
and discuss their properties. 
We first consider sets of black hole solutions 
in $D=5$, 7 and 9 dimensions,
obtained by varying the horizon angular velocity $\Omega$,
while keeping the isotropic horizon radius $r_{\rm H}=1$
and the charge $Q=10$ fixed.

In Fig.~2 (left) we indicate the location of these sets of solutions
within the respective domains of existence,
by exhibiting their scaled angular momentum $J/M^{(D-2)/(D-3)}$ 
versus their scaled charge $Q/M$. 

We exhibit the dependence of the mass $M$ of these solutions
on the horizon angular velocity $\Omega$ in Fig.~2 (right)
and compare with the corresponding $D$-dimensional Myers-Perry solutions,
which have $Q=0$.
For each set of solutions we observe two branches,
extending up to a maximal value of $\Omega$.
The lower branch emerges from the static solution in the limit $\Omega=0$,
on the upper branch the mass diverges in the limit $\Omega \rightarrow 0$.
The maximal value of $\Omega$
depends on the horizon radius $r_{\rm H}$, the charge $Q$, and 
the dimension $D$.
For a fixed value of the charge, its influence 
and thus the deviation from the Myers-Perry
solution decreases with increasing dimension $D$,
as expected from the scaling properties of the solutions \cite{foot1}.

\begin{figure}[h!]
\parbox{\textwidth}
{\centerline{
\mbox{
\epsfysize=10.0cm
\includegraphics[width=70mm,angle=0,keepaspectratio]{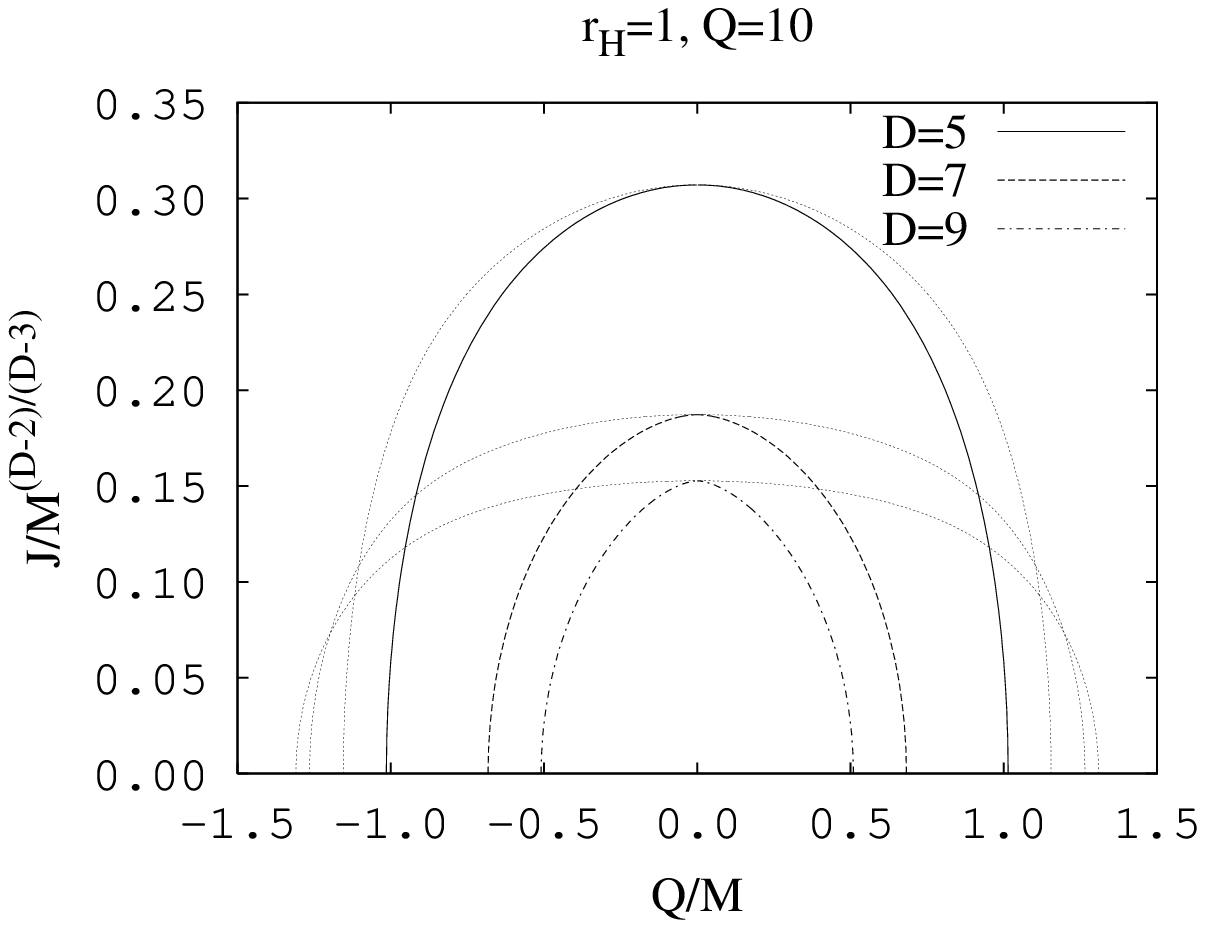}
\includegraphics[width=70mm,angle=0,keepaspectratio]{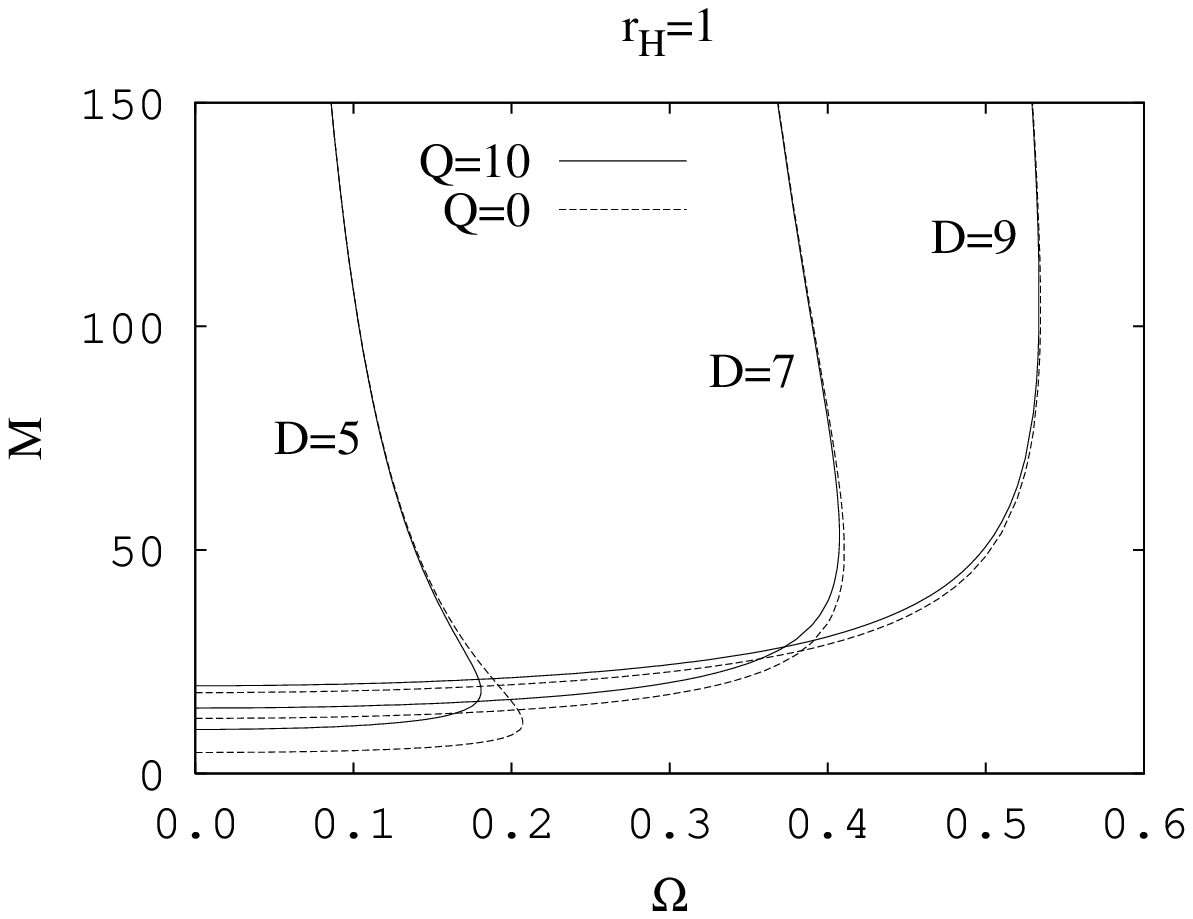}
}}}
\caption{
Sets of non-extremal equal-magnitude angular momenta black hole solutions
in $D=5$, 7 and 9 dimensions,
for fixed isotropic horizon radius $r_{\rm H}=1$,
fixed charge $Q=10$, and varying horizon angular velocity $\Omega$.
Left: Scaled angular momentum $J/M^{(D-2)/(D-3)}$ versus
scaled charge $Q/M$ within the respective domains of existence (represented by
thin dotted lines).
Right: Mass $M$ versus horizon angular velocity $\Omega$ 
(also for Myers-Perry black hole solutions with $Q=0$).
}
\end{figure}

The angular momentum $J$ and the gyromagnetic ratio $g$
are presented in Fig.~3 for the same sets of solutions. 
For small values of the charge $Q$ 
the gyromagnetic ratio has been obtained perturbatively,
with perturbative value $g=D-2$ \cite{aliev}.
We observe, that in the limit $\Omega \rightarrow 0$,
i.e.,~for slow rotation the gyromagnetic ratio
also tends to this perturbative value $g=D-2$.
On the lower branch, which emerges from a static solution,
the gyromagnetic ratio then increases rapidly from its
perturbative value. On the upper branch
the gyromagnetic ratio tends back to its perturbative value
in the limit $\Omega \rightarrow 0$, or whenever the
mass and angular momentum become large so that the
perturbative regime for the charge is reached.

We emphasize that the deviation of the gyromagnetic ratio
from the perturbative value $g=D-2$ is a physical effect.
It is not due to numerical inaccuracy, 
since the error estimate of $10^{-5}$ for $g$ is
typically several orders of magnitude less than the observed deviation 
of $g$ from the perturbative value $g=D-2$.

\begin{figure}[h!]
\parbox{\textwidth}
{\centerline{
\mbox{
\epsfysize=10.0cm
\includegraphics[width=70mm,angle=0,keepaspectratio]{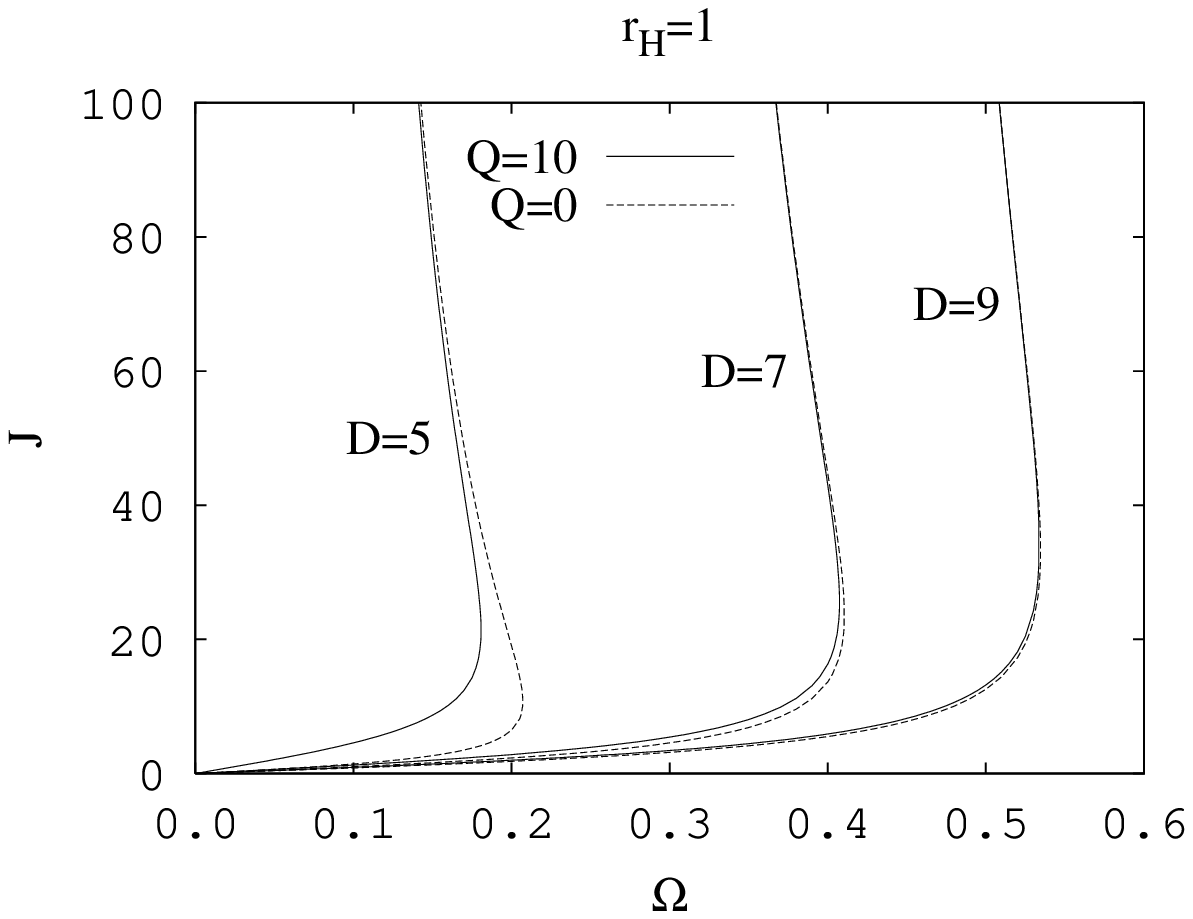}
\includegraphics[width=70mm,angle=0,keepaspectratio]{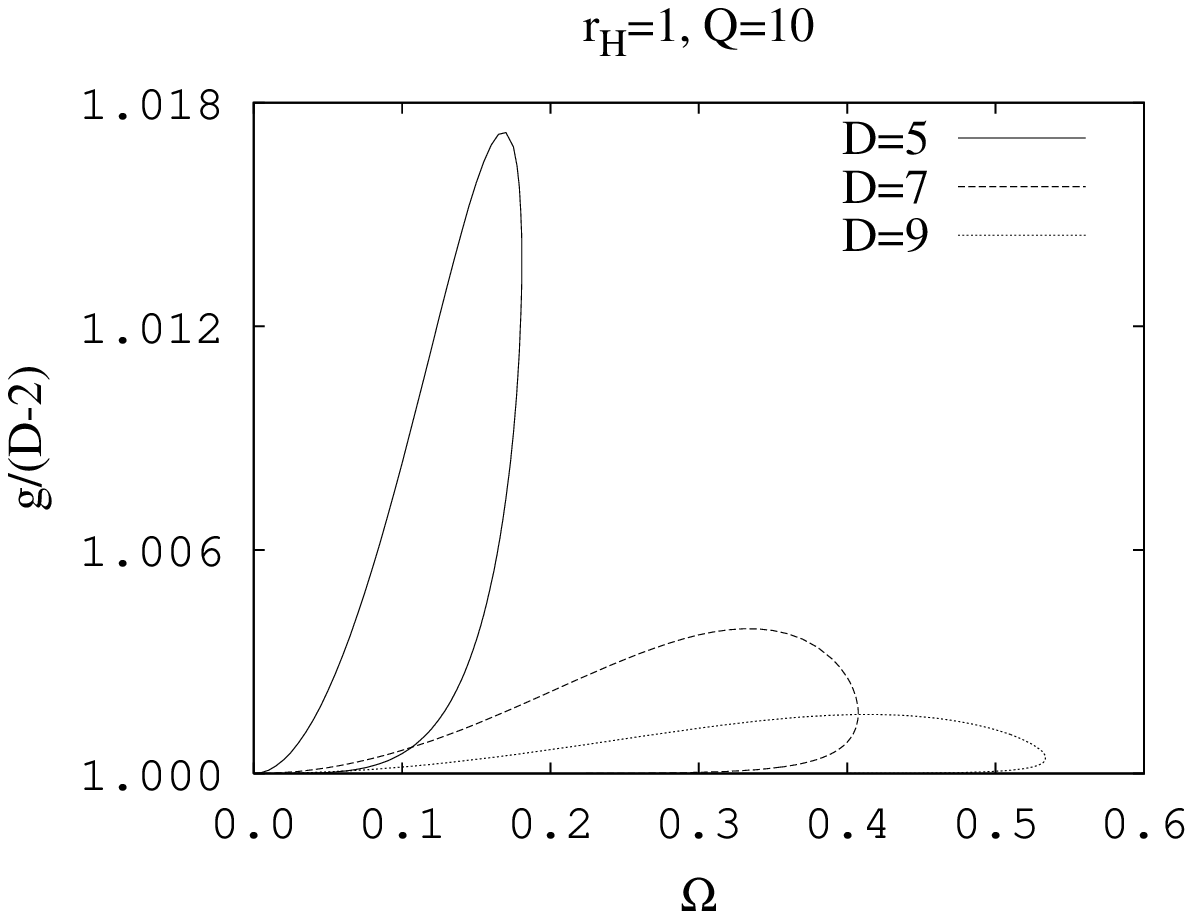}
}}}
\caption{
Same sets of solutions as in Fig.~2.
Left: Angular momentum $J$ versus horizon angular velocity $\Omega$
(also for Myers-Perry black hole solutions with $Q=0$).
Right: Gyromagnetic ratio $g$ versus horizon angular velocity $\Omega$.
}
\end{figure}

\begin{figure}[h!]
\parbox{\textwidth}
{\centerline{
\mbox{
\epsfysize=10.0cm
\includegraphics[width=70mm,angle=0,keepaspectratio]{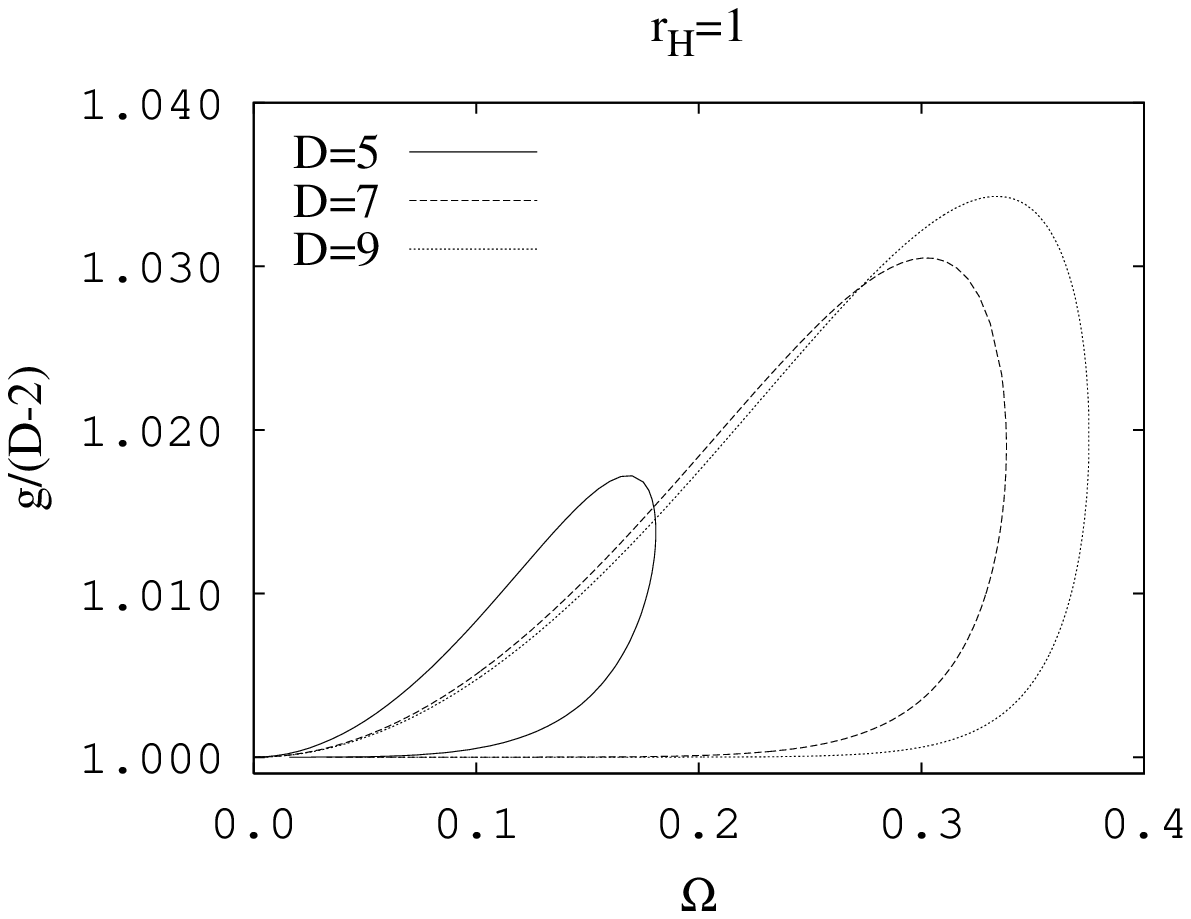}
\includegraphics[width=70mm,angle=0,keepaspectratio]{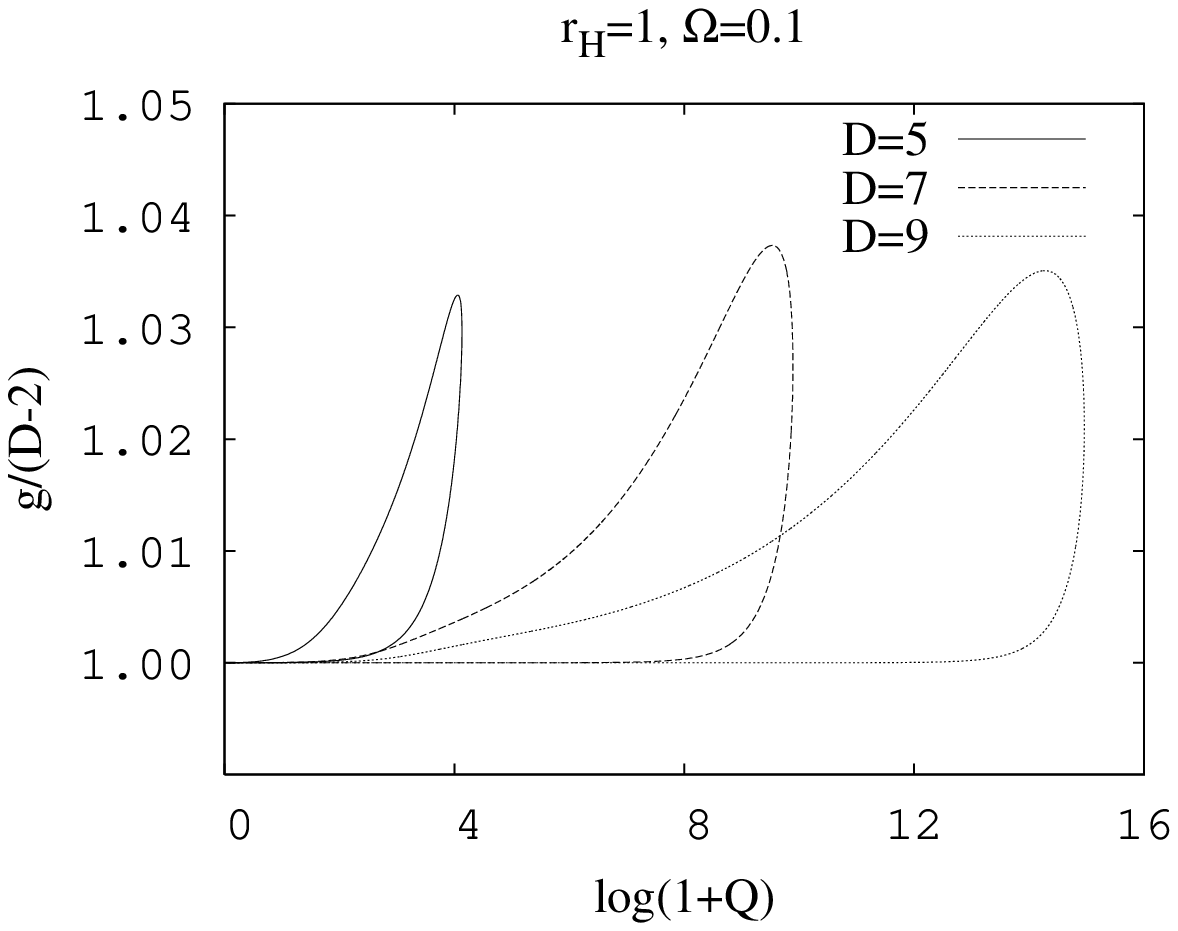}
}}}
\caption{
Left: 
Gyromagnetic ratio $g$ versus horizon angular velocity $\Omega$
for fixed isotropic horizon radius $r_{\rm H}=1$,
fixed charge $Q=10$ for $D=5$, $Q=100$ for $D=7$, $Q=1000$ for $D=9$.
Right:
Gyromagnetic ratio $g$ versus charge $Q$
for fixed isotropic horizon radius $r_{\rm H}=1$,
and fixed horizon angular velocity $\Omega=0.1$
for $D=5$, 7 and 9 dimensions.
}
\end{figure}

As seen in Fig.~3,
the deviation of the gyromagnetic ratio from its respective perturbative value
decreases with increasing dimension $D$ for fixed charge.
To obtain deviations of comparable size for $g$ in higher dimensions,
one has to increase the value of the charge $Q$.
This is illustrated in Fig.~4, where
we compare the gyromagnetic ratio $g$ 
of the previous $D=5$ data set with $Q=10$ to
the $D=7$ and $D=9$ data sets with $Q=100$ and $Q=1000$, respectively.
We also exhibit the gyromagnetic ratio $g$ versus the charge $Q$,
for fixed values of $r_{\rm H}$ and $\Omega$ in Fig.~4. 
Obviously, for small values of $Q$ the gyromagnetic ratio $g$
tends to its respective perturbative value \cite{aliev}.

Horizon properties of the sets of black hole solutions
(presented above in Figs.~2-3) are illustrated in Fig.~5, where we
exhibit their area $A_{\rm H}$ and their surface gravity $\kappa$.
While the horizon area increases monotonically with increasing mass,
the surface gravity tends to zero for large black holes.

\begin{figure}[h!]
\parbox{\textwidth}
{\centerline{
\mbox{
\epsfysize=10.0cm
\includegraphics[width=70mm,angle=0,keepaspectratio]{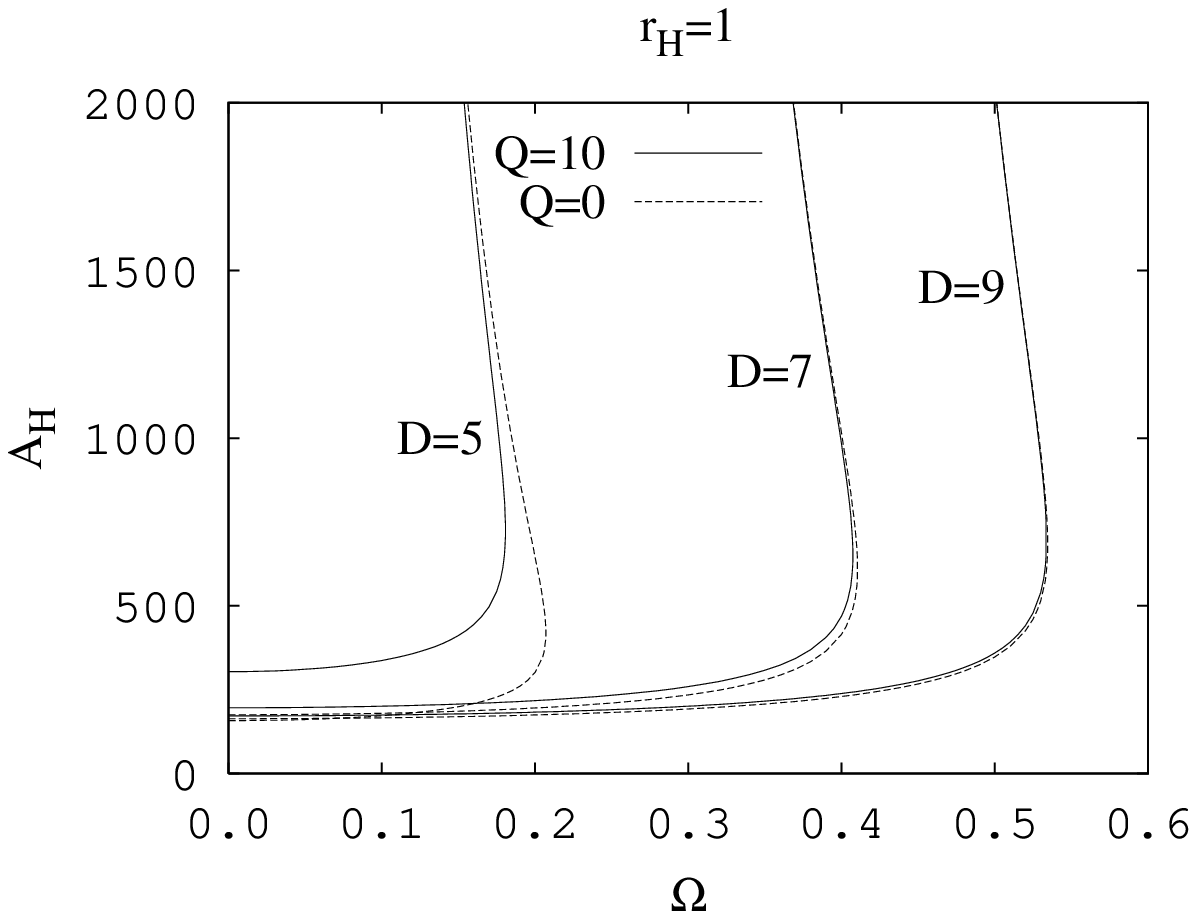}
\includegraphics[width=70mm,angle=0,keepaspectratio]{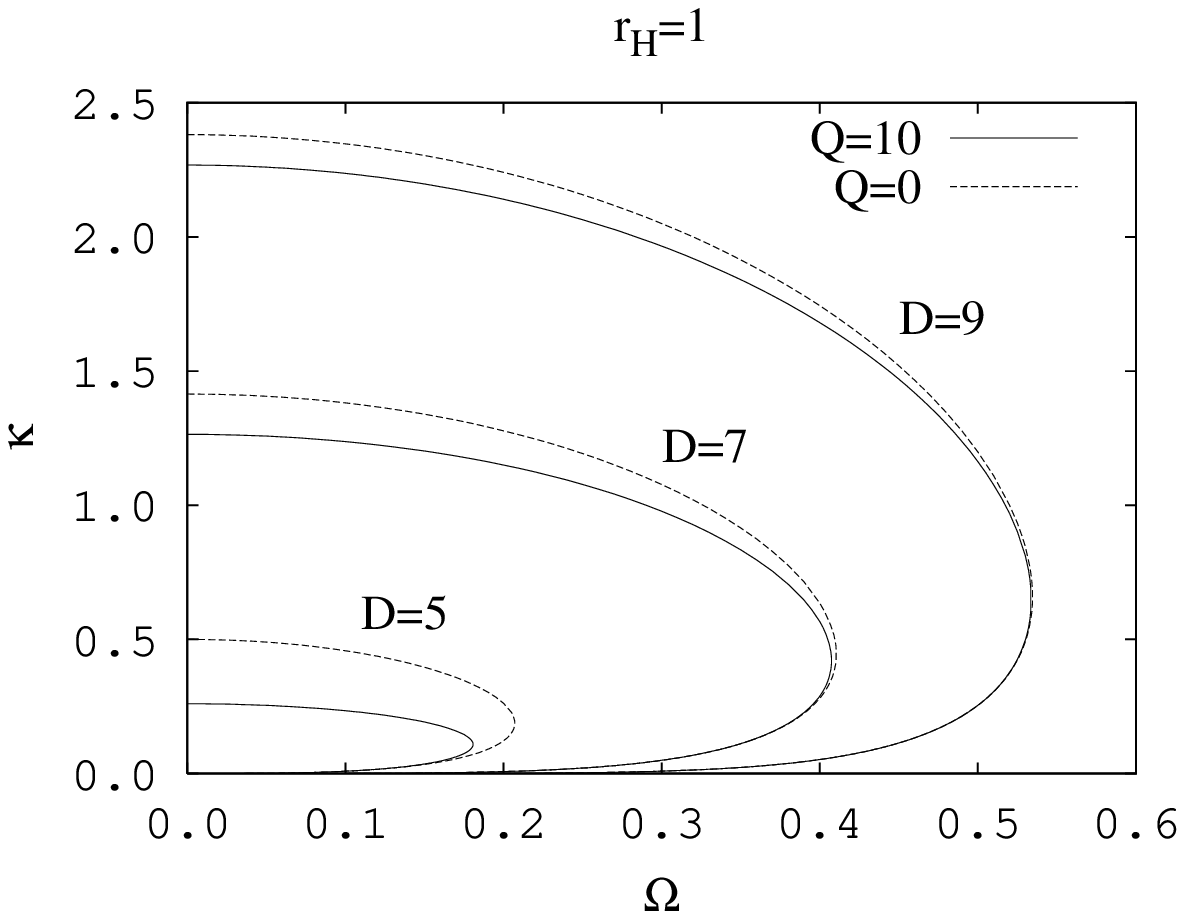}
}}}
\caption{
Same sets of solutions as in Fig.~2 (including the
Myers-Perry black hole solutions with $Q=0$).
Left: Area parameter $A_{\rm H}$ versus horizon angular velocity $\Omega$.
Right: Surface gravity $\kappa$ versus horizon angular velocity $\Omega$.
}
\end{figure}

\section{Conclusions}

We have considered rotating black holes with equal-magnitude
angular momenta in Einstein-Maxwell theory in odd dimensions.
These black holes are asymptotically flat,
and they possess a regular horizon of spherical topology.
We have shown that,
by employing suitable Ans\"atze for the metric and the
gauge potential of these black holes,
the coupled system of Einstein-Maxwell equations
reduces to a set of five ordinary differential equations,
which we have solved numerically.

We have studied the physical properties of these black holes,
in particular their global charges and horizon properties.
The numerical solutions satisfy the generalized Smarr formula
(\ref{smarr}) with high accuracy.
For generic values of the charge and angular momentum
the gyromagnetic ratio of these black holes
differs from $g=D-2$. However, in the limit of
vanishing electric charge or vanishing angular momentum, the gyromagnetic
ratio does tend to the perturbative value $g=D-2$ \cite{aliev}. 
For a fixed value of the charge, its influence on the space-time
decreases with increasing dimension $D$,
as expected from the scaling properties of the solutions \cite{foot1}.

Currently, we are generalizing these results to 
Einstein-Maxwell-Chern-Simons theory,
and to Einstein-Maxwell-dilaton theory.
EMCS black holes, in particular, exhibit already
a number of surprising properties in five dimensions.
These include non-uniqueness, rotational instability, 
or counterrotation \cite{KN}.
Certainly further surprises are waiting here in higher dimensions.

{\bf Acknowledgement}

FNL gratefully acknowledges Ministerio de Educaci\'on y Ciencia for
support under grant EX2005-0078.


\begin{thebibliography}{000}
\bibitem{foot1}
 The solutions have a scaling symmetry
 $\tilde M= \lambda^{D-3} M$,
 $\tilde J= \lambda^{D-2} J$,
 $\tilde Q = \lambda^{D-3} Q$,
 $\tilde r_{\rm H}=\lambda r_{\rm H}$,
 $\tilde \Omega = \Omega/\lambda$,
 $\tilde \kappa = \kappa/\lambda$, etc.
\end{thebibliography}
\end{document}